\documentclass{article}
\usepackage{spconf,amsmath,graphicx,xcolor}
\usepackage{amsmath,amssymb,amsfonts, dsfont}
\usepackage[ruled,vlined]{algorithm2e}
\usepackage[noend]{algpseudocode}

\usepackage{textcomp}

\def\xt{{\mathbf x(t)}}
\def\W{{\mathbf W}}
\def\A{{\mathbf A}}
\def\Wt{{\mathbf W(t)}}
\def\st{{\mathbf s(t)}}
\def\yt{{\mathbf y(t)}}
\def\nt{{\mathbf n(t)}}

\def\xi{{\mathbf x(i)}}

\def\Pt{{\mathbf P(t)}}
\def\htt{{\mathbf h(t)}}
\def\et{{\mathbf e(t)}}
\def\ft{{\mathbf f(t)}}
\usepackage{hyperref}
\hypersetup{
	colorlinks=false,
	linkcolor=blue,
	filecolor=magenta,      
	urlcolor=cyan,
}
\title{Deep-RLS: A Model-Inspired Deep Learning Approach to Nonlinear PCA}

\name{Zahra Esmaeilbeig*, Shahin Khobahi, and Mojtaba Soltanalian\thanks{* Corresponding author (e-mail: zesmae2@uic.edu).}}
\address{Department of Electrical and Computer Engineering, University of Illinois at Chicago, Chicago, USA}

\begin{document}
\maketitle
\begin{abstract}
	In this work, we consider the application of model-based deep learning in nonlinear principal component analysis (PCA).	Inspired by the deep unfolding methodology, we propose a task-based deep learning approach, referred to as  \textit{Deep-RLS}, that  unfolds
	the iterations of the well-known recursive least squares (RLS) algorithm into the layers of a deep neural network in order to perform  nonlinear PCA. In particular, we formulate the nonlinear PCA for the  blind source separation (BSS) problem and show through numerical analysis that \textit{Deep-RLS} results in a significant improvement in the accuracy of recovering the source signals in BSS when compared to the traditional RLS algorithm.
\end{abstract}
\section{Introduction}
\label{sec:intro}
Principal component analysis (PCA)  is a linear orthogonal transformation that determines the eigenvectors or the \emph{basis vectors} associated with the covariance matrix of the observed data. 
The output of the standard PCA is the projection of the data onto the obtained basis vectors, resulting in \textit{principal components} which are mutually uncorrelated. 

As a related problem, blind source separation (BSS)   
~\cite{cardoso1996equivariant}  aims to  recover  statistically independent source signals from  linear mixtures whose composition gains are unknown. In fact,  the only knowledge about the source signals to be recovered in BSS is their independence. This is a stronger assumption than what one requires for PCA, i.e. uncorrelated signal sources. This suggest that performing PCA on such a mixture of  signals  may not successfully separate them, and that higher order statistics is required to verify and ensure the independence of the source signals. 

In light of the above, the notion of \textit{nonlinear correlation} was introduced to verify the independence of signals more effectively~\cite{romano2018unsupervised}. It was also shown that an extension  of PCA to the nonlinear domain (known as \emph{nonlinear PCA}) can separate independent components of the mixture~\cite{romano2018unsupervised}.
In~\cite{cardoso1996equivariant}, an objective function targeting the fourth-order  moments of the  distribution of the  recovered signals is suggested. It is proved that minimizing  this objective function can obtain the  independent  source signals under the  assumption that all sources have negative kurtosis, i.e. they are sub-Gaussian.
Minimizing  such  an objective functions introduces nonlinearities in the signal recovery and the learning process.

In this paper, we propose a deep learning approach to nonlinear PCA that relies on the unfolding of an iterative algorithm into the layers of a deep neural network in order to perform the nonlinear PCA task; i.e. to estimate the source signals for BSS. In particular, we use the classical recursive least squares~\cite{yang1995projection} to construct the network structure and to iteratively estimate the source signals. We experimentally verify the performance of our algorithm and compare it with the traditional RLS algorithm.

\section{Problem Formulation}
\label{sec:Formulation}
\subsection{Nonlinear PCA for Blind Source Separation}
\label{bss}
We begin by considering the longstanding BSS problem in which $m$
statistically independent signals are linearly mixed to yield  $l$ possibly noisy combinations
\begin{equation}
\xt=\A \st+\nt.
\end{equation} 
Let $\xt=[x_1(t), \ldots, x_l(t)]^T$ denote the $l-$dimensional data vector made up of the mixture at time $t$ that is exposed to an additive noise $\nt$.
Given no knowledge of the mixing matrix $\A\in\mathds{R}^{m\times l}$, the goal is to recover the original source  signal vector $\st=[s_1(t), \ldots, s_m(t)]^T$ from the mixture.

A seminal work in this context is~\cite{cardoso1996equivariant} which suggests tuning and updating a separating matrix $\W\in\mathds{R}^{l\times m}$ until the output 
\begin{equation} 
	\yt=\W^T\xt,
\end{equation}
where $\yt=[y_1(t), \ldots, y_m(t)]^T$, is as close as possible to the  source signal vector of interest  $\st$. Assuming there exists a larger number of sensors than the  source signals, i.e. $l\geq m$, we can draw an analogy between  the BSS problem and the task of PCA. In a sense, we are aiming to represent the random vector $\xt$ in a lower dimensional orthonormal subspace, represented by the columns of $\W$, as the orthonormal basis vectors.
By this analogy, both BSS and PCA problems can be reduced to minimizing an objective function of the from:
\begin{equation}
	\label{loss1}
	\mathcal{L} (\W)=\mathbb{E}\left\{\|\xt-\W(\W^T\xt)\|_2^2\right\}.
\end{equation}
Assuming that $\xt$ is a zero-mean vector, it can be shown that the solution to the above  optimization problem is a matrix $\W$ whose columns are the $m$ dominant eigenvectors of  the data covariance matrix $\mathbf{C_x}(t)=\mathbb{E}\left\{ \xt\xt^T \right\}$~\cite{romano2018unsupervised}. Therefore, the principal components or the recovered source signals are mutually uncorrelated. As previously discussed in Section~\ref{sec:intro}, having uncorrelated data is not a sufficient  condition to achieve separation. In other words, the solutions to  PCA and  BSS do not coincide unless we address the  higher order statistics of the output signal $\yt$. By  introducing nonlinearity into~\eqref{loss1},  we  will implicitly target higher order statistics of the signal~\cite{romano2018unsupervised}. This \emph{nonlinear PCA}, which is an extension of the  conventional PCA,  is made possible by considering the signal recovery objective:
\begin{equation}
	\label{nonlinear}
	\mathcal{L}(\W)=\mathbb{E}\left\{\|\xt-\W\mathbf{g}(\W^T\xt)\}\|_2^2\right\},
\end{equation} 
where $\mathbf{g}(\cdot)$ denotes an odd non-linear function applied element-wise on the vector argument. We invite the interested readers to find a proof of the connection between \eqref{nonlinear} and higher order statistics of  the source singals $\st$ in~\cite{pajunen1997least}.

While PCA is a fairly standardized
technique, nonlinear or robust PCA formulations  based on \eqref{nonlinear} tend to be multi-modal with several local optima---so they can be run from various initial points and possibly lead to  different ``solutions"~\cite{karhunen1997class}.
In~\cite{yang1995projection}, a recursive least squares algorithm for subspace estimation is proposed, which is further extended to the nonlinear PCA in~\cite{karhunen1997blind} for  solving the BSS problem. We  consider the algorithm in~\cite{karhunen1997blind} as a baseline for  developing our  deep unfolded  framework for nonlinear PCA.
\subsection{Recursive Least Squares for Nonlinear PCA}
\label{rls}
We consider a real-time and adaptive scenario in which  upon arrival of new data $\xt$, the  subspace of signal at time instant $t$  is recursively updated from the  subspace at time $t-1$ and the new  sample $\xt$~\cite{yang1995projection}. The separating matrix $\W$ introduced in Section~\ref{bss} is therefore replaced by $\Wt$ and updated at each time instant $t$. The adaptive algorithm chosen for this task is the well-known recursive least squares (RLS)~\cite{haykin2014adaptive}.

In the linear case, by replacing the  expectation in~\eqref{loss1} with a weighted sum,
we can attenuate the impact of the older samples, which is reasonable, for instance whenever one deals with a time-varying environment.
In this way, one can make sure the distant past will be forgotten and the resulting algorithm for minimizing~\eqref{loss1} can effectively track the statistical variations of the observed data.
By replacing $\yt=\Wt^T\xt$ and  using an exponential weighting (governed by a \textit{forgetting factor}), the loss function in \eqref{loss1} boils down to:
\begin{equation}
	\label{loss2}
	\mathcal{L}(\Wt)=\sum_{i=1}^{t}\beta^{t-i}||\xt-\Wt \yt||^2,
\end{equation}
with the forgetting factor $\beta$ satisfying $0 \ll\beta \leq1$. Note that $\beta= 1$ yields the
ordinary method of least squares in which all samples are weighed equally while choosing relatively small  $\beta$ makes our estimation rather instantaneous, thus neglecting the past. Therefore, $\beta$ is  usually  chosen to be less than one, but also rather close to one for smooth tracking and filtering.

Note that one may write the 
gradient of the loss function in \eqref{loss2} in its compact form  as 
\begin{equation}
	\label{grad}
	\nabla_{\W} \mathcal{L} (\W)=-2\mathbf{C_{xy}}(t)+2\mathbf{C_{y}}(t)\W,
\end{equation}
where $\mathbf{C_{y}}(t)$ and $\mathbf{C_{xy}}(t)$ are the auto-correlation matrix of $\yt$, 
\begin{equation}
	\label{cy}
	\mathbf{C_{y}}(t) =\sum_{i=1}^{t}\beta^{t-i}\mathbf y(i) \mathbf y(i)^T=\beta \mathbf{C_{y}}(t-1)+\yt\yt^T,
\end{equation}
and the cross-correlation matrix  of $\xt$ and $\yt$, 
\begin{equation}
	\mathbf{C_{xy}}(t) =\sum_{i=1}^{t}\beta^{t-i}\mathbf x(i) \mathbf y(i)^T=\beta \mathbf{C_{xy}}(t-1)+\xt \yt ^T,
\end{equation}
at the time instance $t$, respectively. Setting the gradient~\eqref{grad} to zero will result in the close-form separating matrix,
\begin{equation}
	\label{sol}	
	\Wt=\mathbf{C_{y}}^{-1}(t)\mathbf{C_{xy}}(t).
\end{equation}

\begin{algorithm}[t]
	\caption{RLS Algorithm for Performing PCA}
	\begin{algorithmic}[1]
		\State\textbf{Initialize} $W(0)$ and $P(0)$
		
		\State\For {$t=0,1,\ldots, T$}{}
		\State{\indent} $\yt=\mathbf{W}^T(t-1)\xt$
		\State{\indent}	$\htt=\mathbf{P}(t-1)\yt$
		\State{\indent}	$\ft=\frac{\htt}{\beta+\mathbf y(t)^T \htt}$
		\State{\indent}	$\Pt=\beta^{-1}[\mathbf{P}(t-1)-\ft \mathbf h(t)^T]$
		\State{\indent}	$\et=\xt-\mathbf{W}(t-1)\yt$
		\State{\indent}	$\Wt=\mathbf{W}(t-1)+\et \mathbf f (t)^T$
	\end{algorithmic}
	\label{algo1}
\end{algorithm}

A recursive computation of $\Wt$ can be achieved using the RLS algorithm~\cite{haykin2014adaptive}. In RLS, the matrix inversion lemma enables a recursive computation of $\Pt=\mathbf{C_{y}}^{-1}(t)$; see the derivations in Appendix. At each iteration of the RLS algorithm, 
$\Pt$ is recursively computed as
\begin{equation}\label{rec}	
	\Pt=\beta^{-1}\mathbf{P}(t-1)-
	\frac{\beta^{-2}\mathbf{P}(t-1)\yt \yt^T\mathbf{P}(t-1)}
	{1+\beta^{-1}\yt^T \mathbf{P}(t-1)\yt}.
\end{equation}
Consequently, the RLS algorithm provides the estimate $\yt$ of the source signals. The steps of the RLS algorithm are summarized in Algorithm~\ref{algo1}.

It appears that extending the application of RLS to the  nonlinear PCA  loss function  in \eqref{nonlinear} is rather straightforward. To accomplish this task, solely step 3 of Algorithm~\ref{algo1} should be modified to $\yt=g(\W^T(t-1)\xt)$ in order to meet the nonlinear PCA criterion~\cite{yang1995projection}. In the following, we \emph{unfold} the iterations of the modified Algorithm~\ref{algo1}, for nonlinear PCA  onto the layers of a deep neural network where each layer resembles 
one iteration of the RLS algorithm. Interestingly,
one can fix the complexity budget of the inference framework via
fixing the number of layers, and apply the RLS-based
proposed method to yield an estimation of the source signals.
\section{Deep Unfolded RLS for Nonlinear PCA}
Deep neural networks (DNN) are one of the most studied approaches in machine learning owing to their significant potential. They are usually used as a black-box without incorporating any knowledge of the system model. Moreover, they are not always practical to use as they require a large amount of training data and considerable computational resources. Hershey, et al., in~\cite{hershey2014deep}, introduced a technique referred to as \textit{deep unfolding} or \textit{unrolling}  in order to enable the community to address the mentioned issues with DNNs. The deep unfolding technique lays the ground for bridging the gap between well-established iterative signal processing algorithms that are model-based in nature and  deep neural networks that are purely data-driven. 
Specifically, in deep unfolding, each layer of the DNN is designed to resemble one
iteration of the original algorithm of interest. Passing the signals through such a deep network is in essence similar to executing
the iterative algorithm a finite number of times, determined by the number of layers. In addition,
the algorithm parameters (such as the model parameters and
forgetting parameter in the RLS algorithm) will be reflected in the parameters and weights of the constructed DNN. The data-driven nature of the emerging deep network thus enables improvements over the original algorithm.  Note that the constructed network may be trained using back-propagation, resulting
in model parameters that are learned from the real-world training
datasets. In this way, the trained network can be naturally
interpreted as a parameter optimized algorithm, effectively
overcoming the lack of interpretability in most conventional
neural networks~\cite{monga2019algorithm}. In comparison with a generic DNN, the unfolded network has much fewer parameters and therefore requires a more modest size of training data and  computational resources. The deep unfolding  technique has been deployed for many signal processing problems and has dramatically improved the convergence rate of the state-of-the-art model-based iterative algorithms; see, e.g.,~\cite{khobahi2019deep,khobahi2020deep,solomon2019deep} and the references therein.

Due to  the promises of deep unfolding in addressing the shortcomings of both generic deep learning methods  and model-based signal processing algorithms, we have been motivated to  deploy this technique to improve the recursive least squares solution for nonlinear PCA. As shown in~\cite{karhunen1997blind}, when applied to a linear mixture of  source signals (i.e., the BSS problem), the RLS algorithm usually approximates the true source signals well and successfully separates them. However, the number of iterations needed to converge may vary greatly depending on the initial values and the forgetting parameter $\beta$. Inspired by the deep unfolding technique, 
we introduce \textit{Deep-RLS}, our deep learning-based 
framework which is designed based on the modified iterations of the algorithm~\ref{algo1}. More precisely, the  dynamics of the
$k$-th layer of \textit{Deep-RLS} are given as: 
\begin{subequations}
	\label{deeprls}
	\begin{align}
		\mathbf y(k)& =\mathbf g(\mathbf{H}_{k}\mathbf W^T(k-1)\mathbf x(k)+\mathbf{b}_{k}	), \label{deeprls1}\\
		\mathbf h(k)&=\mathbf P(k-1)\mathbf y(k), \label{deeprls3}	\\
		\mathbf f(k)&=\frac{\mathbf h(k)}{\mathbf \omega_{k}+\mathbf y(k)^T\mathbf h(k)},\label{deeprls2} \\
		\mathbf P(k)&=\omega_{k}^{-1}[\mathbf P(k-1)-\mathbf{f}(k)\mathbf{h}(k)^T],\label{deeprls4}\\
		\mathbf e(k)&=\mathbf{x}(k)-\mathbf{H}_{k}\W(k-1)\mathbf y(k), \label{deeprls5}	\\
		\W(k)&=\mathbf{H}_{k}\W(k-1)+\mathbf e(k)\mathbf f(k)^T,\label{deeprls6} 
	\end{align}
\end{subequations}
where $\mathbf x(k)$ is the  data vector at the time instance $k$, $\mathbf g(.)$ is a nonlinear activation function which can be chosen by considering the distribution of the source signals, $\omega_k\in\mathds{R}$ represents the trainable forgetting parameter, and $\mathbf{H}_{k}\in\mathds{R}^{m\times m}$ and $\mathbf{b}_{k}\in\mathds{R}^{m}$ denotes the set of trainable weights and biases of the $k$-th layer, respectively. In~\cite{karhunen1997class}, it was shown that for source signals $\st$ with sub-Gaussian distribution, $\mathbf g(\mathbf x)=tanh(\mathbf x)$ results in convergence of the nonlinear PCA to the true source signals.
\begin{algorithm}[t]
	\caption{Training Procedure for \textit{Deep-RLS}}
	\begin{algorithmic}[1]
		\State \textbf{Initialize} $\W(0)$ and $\mathbf P(0)$
		\State \For {$epoch=1, \ldots, N$}{\For {$k=1, \ldots, T$}
			{Feed $\mathbf x(k)$ to $k$-th layer of the network \newline
				\indent Apply the recursion in~\eqref{deeprls}} 
			
			Compute the loss function~\eqref{eq1} 
			
			\indent Use backpropagation to update $\{\Gamma_k\}_{k=1}^T$ }
	\end{algorithmic}
	\label{algo2}
\end{algorithm}

Given $T$ samples of the data vector $\xt$,
our goal is to optimize the parameters $\{\Gamma_k\}_{k=1}^{T}$ of the DNN, where 
\begin{equation}
\Gamma_k= \{\mathbf{H}_{k},\mathbf{\omega}_{k},\mathbf b_{k}\}.
\end{equation}
 The output of the $k$-th layer is an approximation of the  source signals at the time instance $k$. 
 For training of  the proposed \textit{Deep-RLS} network, we  consider cumulative  MSE loss of the layers. In designing the training procedure, one needs to consider the constraint that the forgetting parameter must satisfy $0<
\beta\leq 1$. Hence, in order to impose such a constraint, one can regularize the loss function ensuring that the network chooses proper weights $\{\omega_{k}\}_{k=1}^T$ corresponding to a feasible forgetting parameter at each layer~\cite{khobahi2019model}. Accordingly, we define the loss function used for the training of the proposed architecture as follows:
\begin{equation} \label{eq1}
	\begin{split}
		\mathcal{L}(\W(k)) & =\underbrace{\sum_{k=1}^{T}||\mathbf x(k)-\W(k)\mathbf y(k)||^2}_\text{accumulated  loss of all layers}+\\
		& \underbrace{	\lambda \sum_{k=1}^{T}
			\mathrm{ReLU}(-\mathbf{\omega}_{k})
			+\lambda \sum_{k=1}^{T} \mathrm{ReLU}(\mathbf{\omega}_{k}-1) }_\text{regularization term for the forgetting parameter},
	\end{split}
\end{equation}
where  $\mathrm{ReLU}(\cdot)$ is  the well-known Rectifier
Linear Unit function extensively used in the deep learning literature.
We have presented the employed training  process in Algorithm~\ref{algo2}.

\section{Numerical Result}
In this section, we demonstrate the performance of the proposed \textit{Deep-RLS} for nonlinear PCA in the case of blind source separation. The proposed framework was implemented using the PytTorch library~\cite{NEURIPS2019_9015} and the Adam stochastic optimizer~\cite{kingma2014adam} with a learning rate of $10^{-3}$ for training purposes.

\begin{figure}[t]
	
	\centering 
	\includegraphics[width=0.45\textwidth]{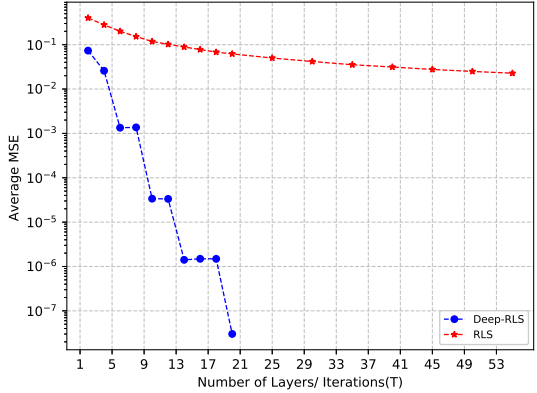}
	\caption{The average MSE of recovering $m=2$ source signals using the \textit{Deep-RLS} network  vs. the number of layers/iterations $T$ when trained for $N=50$ epochs. The proposed method significantly outperforms the RLS algorithm with $\beta=0.99$.}
	\label{fig11}
\end{figure}
\begin{figure}[t]
	
	\centering 
	\includegraphics[width=0.45\textwidth]{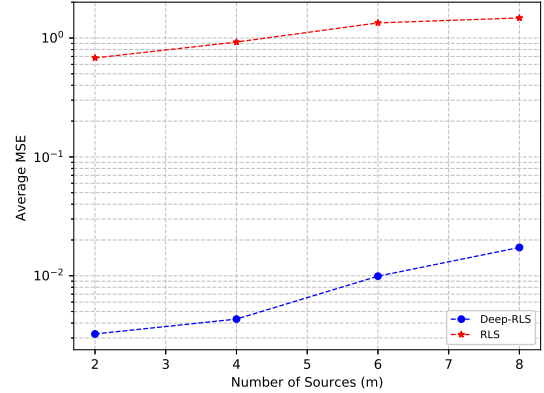}
	\caption{The performance of \textit{Deep-RLS} and the the traditional RLS algorithm when applied to the recovery of a growing number of  source signals.}
	\label{fig2}
\end{figure}

The training was performed based on the data generated
via the following model. For the time interval $t=0,1,...,T$, each element of the vector
$s(t)$ is to be generated from a sub-Gaussian distribution. For data generation purposes, we have assumed the source signals to be i.i.d. and uniformly distributed, i.e.,
$s(t)\sim \mathcal{U}(0,1)$. The mixing matrix $\mathbf{A}$ is assumed to be
fixed and generated once according to a Normal distribution, i.e.,
$\mathbf A \sim \mathcal{N}(\mathbf 0, \mathbf I)$.
We performed  the training of the proposed Deep-RLS using the batch learning process with a batch size of 40 and trained the network for $N=50$ epochs. A training set of size $10^3$ and test set of size $10^2$ was chosen. We used the average of the mean-square-error (MSE), 
$\mathrm{MSE} =(1/T)\sum_{k=1}^{T}||\mathbf s(k)-\mathbf y(k)||^2_2$, as the performance metric. In Fig.~\ref{fig11}, the performance of  RLS  implemented with $\beta=0.99$ at different number of iterations is compared with the \textit{Deep-RLS} network trained with the same number of layers.  It can be observed that the average MSE obtained using the proposed architecture is much lower than RLS, even with fewer number of layers/iterations.
Fig.~\ref{fig2} demonstrates the performance of the proposed
\textit{Deep-RLS} network and the original RLS algorithm in terms of the average MSE for a growing
number of source signals $m$. It can be observed from both Figs. 1 and 2 that the proposed method achieves a far superior performance than that of the original RLS algorithm.

\section{Conclusion}
We considered the application of model-based machine
learning, and specifically the deep unfolding technique, for nonlinear PCA. A deep network based on the well-known RLS algorithm, which we call \textit{Deep-RLS},  was proposed that outperforms its traditional model-based counterparts.

\appendix
\section{Appendix: The RLS Recursive Formula}
\label{sec:app}
Let $\mathbf A$, $\mathbf B$, and $\mathbf D$ be positive definite matrices so that
$
 	\mathbf A=\mathbf B^{-1}+\mathbf{cD}^{-1}\mathbf{c}^T.
$ 
Using the matrix inversion lemma, the inverse of $\mathbf A$ can be expressed as
\begin{equation}
	\mathbf A^{-1}=\mathbf B-\mathbf{Bc}(\mathbf{D}+\mathbf{c}^T\mathbf{Bc})^{-1}\mathbf c^T\mathbf B. 
\end{equation}
Now, assuming  that the auto-correlation matrix $\mathbf{C_y}(t)$ is  positive definite (and thus nonsingular), by choosing $\mathbf{A=C_y}(t)$, $\mathbf{B}^{-1}=\beta\mathbf{C_y}(t-1)$,$\mathbf{c}=\yt, \mathbf D^{-1}=1$, one  can compute $\Pt=\mathbf{C_y}^{-1}(t)$ as proposed in \eqref{rec}.
\bibliographystyle{IEEEbib}
\bibliography{refs}

\end{document}